\documentclass[english,showpacs,twocolumn]{revtex4-1}
\usepackage{mathptmx}
\usepackage[T1]{fontenc}
\usepackage[latin9]{inputenc}
\usepackage{color}
\usepackage{amsmath}
\usepackage{amssymb}
\usepackage{hyperref}
\usepackage{graphics}
\usepackage{graphicx}
\usepackage{epsfig}
\usepackage{bm}
\usepackage{epstopdf}
\usepackage{mathtools}
\usepackage{bm}
\newcommand{\uvec}[1]{\boldsymbol{\hat{\textbf{#1}}}}
\usepackage{subcaption}
\usepackage{lipsum}
\usepackage{mwe}
\usepackage{adjustbox}
\graphicspath {{figure/}}
\usepackage{tikz}
\usepackage{pgfplots}
\usepackage{xmpmulti}

\usepackage{babel}
\begin{document}
\title{
Dust vortex flow analysis in weakly magnetized plasma
}
\author{Prince Kumar}
\email{kumarprincephysics@gmail.com}
\author{Devendra Sharma}
\email{devendra@ipr.res.in}
\affiliation{Institute for Plasma Research, HBNI, Bhat, Gandhinagar - 382 428, Gujarat, India, and}
\affiliation{Homi Bhabha National Institute, Mumbai, Maharashtra-400 094, India}
\date{\today}
\begin{abstract}
Analysis of driven dust vortex flow is presented in a weakly magnetized plasma.
The 2D hydrodynamic model is applied to the confined dust cloud in a 
non-uniform magnetic field in order to recover the dust vortex flow driven in a 
conservative force field setup, in absence of any non-conservative fields 
or dust charge variation.
Although the time independent electric and magnetic
fields included in the analysis provide conservative forcing mechanisms, 
when the a drift based mechanism, recently observed in a dusty plasma 
experiment by [M. Puttscher and A. Melzer, Physics of Plasmas, 21,123704(2014)] is considered, the dust vortex flow solutions 
are shown to be recovered.
We have examined the case where purely ambipolar electric field, generated by 
polarization produced by electron $\textbf{E}\times\textbf{B}$ drift, 
drives the dust flow. A sheared $\textbf{E}\times\textbf{B}$ drift flow 
is facilitated by the magnetic field gradient, driving the vortex flow in 
the absence of ion drag. The analytical stream-function solutions have been 
analyzed with varying magnetic field strength, its gradient and kinematic 
viscosity of the dust fluid. The effect of $\textbf{B}$ field gradient is 
analyzed which contrasts that of $\textbf{E}$ field gradient present in the 
plasma sheath.

\end{abstract}
\maketitle
\section{Introduction}\label{introduction} 
Vortex flow in a charged fluid are highly relevant to generation of magnetic
fields in nature and equilibrium configurations of magnetic confinement 
plasma experiments {\color{blue}{\cite{saitou}}}. Quasi neutral electron-ion plasmas with highly charged 
dust particles present as third species {\color{blue}{\cite{shukla2001}}}, 
or a dusty plasma, provides a 
setup where vortex flow of the charged dust fluid is often present  {\color{blue}{\cite{shukla}}}  and can 
be studied at very accessible spatio-temporal scale. The dust vortex flow in 
plasmas is modeled using the macroscopic 2D hydrodynamic formulation in 
magnetized plasma. The effects of an ambient magnetic field are expected 
to be moderate on the dust as long as the magnetic field is not strong enough 
to magnetize the dust particles. Recent experimental studies have however 
shown that in a weakly magnetized plasma where only electrons are magnetized,
dust motion can have finite effects of the magnetic field via magnetization 
of electrons {\color{blue}{\cite{Nunomura,Cheung,Konopka,Karasev}}}. This paper presents a hydrodynamic formulation for the dust 
vortex flow accounting for effects of weak magnetization as observed and 
quantified in these recent experiments. With the availability of advanced 
magnetized dusty plasma experiments like MDPX {\color{blue}{\cite{hall,Thomas}}}  the steady of collective 
dust dynamics, described here in weak to strongly magnetized plasma regime, 
may be possible with greater flexibility.

The existing dusty plasma studies show that the dust species in a plasma is 
subjected to various forces. The effects of forces on dust due to
ion drag {\color{blue}{\cite{Khrapak1,Khrapak}}}, neutral drag and 
electrostatic forces, has been extensively studied both experimentally and 
theoretically in literature. Dust Vortex flow structures which are driven 
by non-conservative force fields, like ion drag force 
{\color{blue}{\cite{M.Kaur}}}, neutral flow
{\color{blue}{\cite{Vladimirov,S.Mitic,M.Schwabe}}} have been observed in 
experiments. 
Dust Vortex flow have also been observed under external forces 
{\color{blue}{\cite{Law,Uchida,Klindworth,Miksch,saitou}}}. 
Recently, rotating dust structures have also been observed in weakly 
magnetized plasmas {\color{blue}{\cite{Nunomura,Cheung,Konopka,Karasev}}} 
where dust dynamics is again interpreted as governed by the non-conservative 
forces like, ion drag and neutrals flow. 

In laboratory experiment, dynamics of both para magnetic and diamagnetic 
(Melamine-formaldehyde or MF) particles was investigated in the presence 
of gradient in magnetic field by Puttscher and Melzer {\color{blue}{\cite{Melzer2}}},
finding that only para-magnetic particles responded to magnetic field 
gradient. An interesting dynamics of diamagnetic particles was however 
also reported by Puttscher and Melzer {\color{blue}{\cite{Melzer}}} which is governed 
by an ambipolar electric field generated due to magnetized drifting electrons.
Since the charged dust particles respond directly to an electrostatic field,
their motion is governed by a conservative field which in usual cases does 
not produce a vortex flow, unlike a nonuniform drag or frictional force {\color{blue}{\cite{Khrapak1,Khrapak}}}. 
In this work we analyze the dust flow in a weakly magnetized setup with 
a similar ambipolar forcing field and recover the dust vortex flow when the 
magnetic field has finite gradient.

Melzer and Puttscher {\color{blue}{\cite{Melzer3}}} presented a force
balance mechanism that can explain the motion of dust particles in the 
presence of weak homogeneous magnetic field. Since only electrons were 
magnetized and drifted in $\textbf{E}\times\textbf{B}$ direction, they produced 
an ambipolar electric field that acted both on the ions and the dust particles.
They observed that although at low gas pressure, motion of dust particle was 
driven by the ion drag which acted flow along $\textbf{E}\times\textbf{B}$, 
at sufficiently high gas pressure the overall dust dynamics was governed 
purely by the ambipolar electric field and they moved against the 
$\textbf{E}\times\textbf{B}$ direction.

The $\textbf{E}\times\textbf{B}$ effect on dust particles observed by 
Melzer and Puttscher {\color{blue}{\cite{Melzer3}}} arises because of a 
sheath electric field. Since this may be strongly sheared in the sheath 
region, it motivates the idea as to weather a sheared 
$\textbf{E}\times\textbf{B}$ drift can drive a vortex motion of a suspended 
dust fluid.
Considering this, we study driven flow field of confined dust 
fluid which is suspended in the plasma sheath in the presence of a non-uniform 
weak magnetic field. Our analysis shows that the ambipolar electric field
can act as a source of finite vorticity in the dust flow dynamics. We derive 
and use the circulation of the ambipolar field generated by 
$\textbf{E}\times\textbf{B}$ drift of the electrons as a driver for
vortex flow of the dust motion and study its behavior in presence of 
nonuniform magnetic field. The results show dependence of intensity of 
dust vortex motion on the strength of magnetic field and its gradient.

The paper is organized as follows. The 2D hydrodynamic model for confined 
dust fluid, 
in Cartesian geometry, with non-uniform magnetic field is introduced in 
sec.{\color{blue}{\eqref{NM}}}. In sec. {\color{blue}{\eqref{NM1}}} 
a boundary value problem constructed in Cartesian geometry for dust 
stream-function. In order to find analytic dust stream-function solutions, 
the boundary value problem is converted into an eigenvalue problem in which 
the dust stream-function and the driver are expressed in terms of 
suitable eigenfunctions. 
Dust stream-function solution is analyzed in sec. 
{\color{blue}{\eqref{NM2}}} with variation in applied non-uniform magnetic 
field and kinematic viscosity $\mu$ of dust fluid. Dust vortex solutions for 
multipolar form of the ambipolar field are analyzed in 
sec.{\color{blue}{\eqref{MSD}}}. The summary and conclusion of 
the result has been presented in sec.{\color{blue}{\eqref{SC}}}.
\section{THE DUST VORTEX MODEL IN A WEAKLY MAGNETIZED PLASMA}\label{NM}
\begin{figure}[htb]
\setkeys{Gin}{width=\linewidth}
        \includegraphics{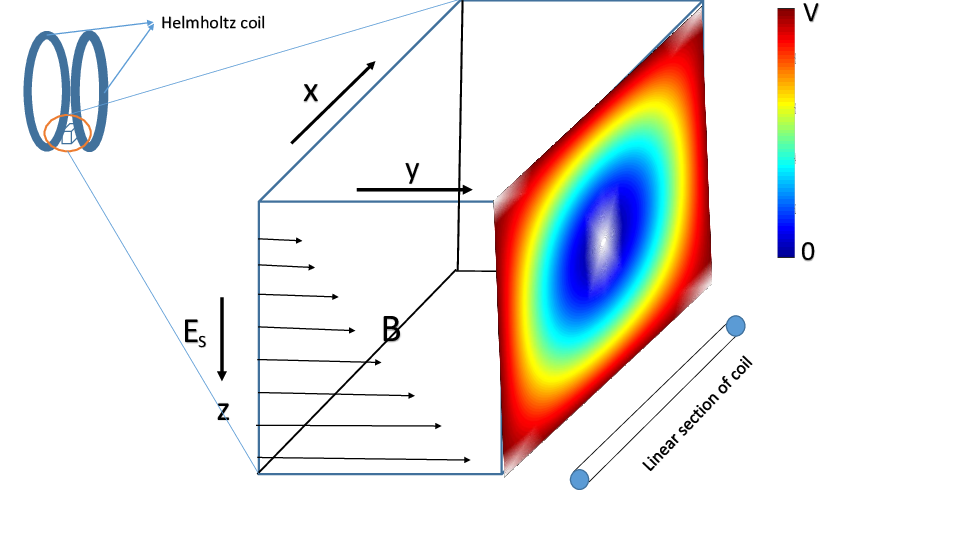}
    \caption{Schematic of the set up in Cartesian geometry with the contour plot showing the strength of effective confining potential V(z,x) for the dust fluid.}
    \label{fig:bear}
\end{figure}

The setup of confined dust fluid considered here is motivated by the 
experiment by Puttscher and Melzer {\color{blue}{\cite{Melzer}}}, 
who studied the behavior of dust particle motion in mutually perpendicular 
electric and magnetic field in the sheath region of an rf discharge. 
We consider a dust cloud modeled as a fluid suspended in plasma where 
both electrostatic and gravitational fields acting on it are mutually balanced.
A nonuniform magnetic field is considered with a variation approximated as 
linear over a relatively small dimension of the dust confinement region
as compared to scale lengths of the variation.
In a Cartesian setup as described in Fig.~\ref{fig:bear}, we accordingly 
use a magnetic field aligned with $y$ axis with its variation along $z$ axis,
$\textbf{B}(z) = B_0(1+\alpha z)\uvec{y}$,
produced locally, for example, by a section of a coil directed along $x$ 
while coil axis directed along y-direction. A constant sheath electric field 
$\textbf{{E}}$ = $E_s\uvec{z}$ is considered present 
in the z-direction.

We study the 2-dimension dust fluid dynamics in a \:$x-z$\: plane in a dust 
confinement domain ranging in the limits,
\:$0 < x/L_x <1$\ and  \:$0 < z/L_z <1$, respectively; assuming symmetry 
along $y$.
The basic hydrodynamics equations include the $z$ and $x$ components of 
the Navier-Stokes equation,

\begin{equation} \label{1}
 \begin{split}
 \frac{\partial U_z}{\partial t} + U_z\frac{\partial U_z}{\partial z}+U_x\frac{\partial U_z}{\partial x} =-{\frac{1}{\rho}}\frac{\partial P}{\partial z} - \frac{\partial V}{\partial z} + \mu  \nabla^2U_z +	\frac{q_d}{m_d} E_a\\-\nu(U_z - W_z),
\end{split}
\end{equation}

\begin{equation} \label{2}
 \begin{split}
 \frac{\partial U_x}{\partial t} + U_z\frac{\partial U_x}{\partial z}+U_x\frac{\partial U_x}{\partial x} =-{\frac{1}{\rho}}\frac{\partial P}{\partial x} - \frac{\partial V}{\partial x} + \mu  \nabla^2U_x + \frac{q_d}{m_d} E_a\\-\nu(U_x - W_x),
\end{split}
\end{equation}

respectively, and equation of continuity for the incompressible dust fluid, 
$\nabla\cdot\ \textbf{U}=0$, written 
as,

\begin{equation} \label{continuity}
\frac{\partial U_z}{\partial z}+\frac{\partial U_x}{\partial x} = 0,
 \end{equation}

where V is the effective confining potential, $q_d$ and $m_d$ are the dust charge and mass, respectively,
$\textbf{U}$ is the dust velocity and $\textbf{W}$ is the flow velocity 
of the neutral fluid. \:$P$\ and \:{$\rho$}\ are the pressure and mass 
density of the dust fluid, respectively,\:{$\nu$}\ is coefficient of the friction with the neutral fluid acting on the dust and \:{$\mu$\ is its kinematic 
viscosity. 
The ion drag is ignored considering the limit of high pressure
{\color{blue}{\cite{Melzer}}} where the dust dynamics is mainly 
governed by the ambipolar field $\textbf{E}_\textbf{a}$ due to 
electron fluid drifting past ions 
with the $E\times B$ drift 
perpendicular to the plane containing $\textbf{E}$ and $\textbf{B}$. 

 We estimate $\textbf{E}_\textbf{a}$ from the current 
density $\textbf{j}$ of the drifting  electrons. 
We begin by considering the electron momentum balance for time independent
condition in presence of resistivity $\eta$,
\begin{eqnarray}
0=\textbf{E}_{T}+\textbf{v}_{e}\times \textbf{B}-\eta\textbf{j},
\label{e-momentum-general}
\end{eqnarray}
where $\textbf{E}_{T}=\textbf{E}+\textbf{E}_{a}$ and
\begin{eqnarray}\label{def-j}
\textbf{j} = n_e q_e \textbf{v}_\textbf{e},
\end{eqnarray}
where $q_e$ is the elementary charge on electron and 
$n_{e}$ is electron density.
As mentioned above, $\textbf{E}=E_{s}\hat{z}$, and $\textbf{B}=B\hat{y}$ 
are externally applied fields.
In the absence of resistivity $\eta \textbf{j}=\textbf{E}_{a}=0$ and the 
lowest order expression for the force balance, containing terms directed 
purely along the $z$-axis, is recovered,
\begin{eqnarray}
0=\textbf{E}+{\textbf{v}_{e}\times \textbf{B}},
\label{e-momentum-pure}
\end{eqnarray}
which yields the $E\times B$ velocity directed along $-\hat{x}$,
\begin{equation} \label{ExB}
\textbf{v}_\textbf{e}=\frac{\textbf{E}\times\textbf{B} }{|\textbf{B}|^2}.
\end{equation}
When reisitivity $\eta$ is finite, i.e., electrons lose momentum via collisions 
while drifting along $x$, the frictional force $-\eta \textbf{j}$ must be 
balanced by a correction in the electric field (or in the effective drift) as 
required in the equilibrium state, 
such that the general force equilibrium (\ref{e-momentum-general}) emerges. 
For the case $E_{a}\ll E$, when the lowest order equilibrium 
(\ref{e-momentum-pure}) can be subtracted from 
(\ref{e-momentum-general}), the residual force balance, predominantly 
along $x$-axis, reads,
\begin{eqnarray}\label{E-ambipolar}
0=\textbf{E}_{a}-\eta\textbf{j}.
\end{eqnarray}
Thus $\textbf{E}_{a}$ can be estimated if the lowest order $E\times B$ drift 
expression (\ref{ExB}) is used to determine the current density $\textbf{j}$
given by (\ref{def-j}) which is then substituted in Eq.~(\ref{E-ambipolar}), 
obtaining, 
\begin{equation} \label{ambiploar}
\textbf{E}_\textbf{a} \approx \eta n_e q_e \frac{\textbf{E}\times\textbf{B} }{|\textbf{B}|^2}
\end{equation}
The resistivity $\eta$ has contributions both from electron-ion and 
electron-neutral collisions. For simplicity, however, in the present 
treatment we consider $\eta$ to 
be the transverse Spitzer resistivity of the plasma.

For the dust flow which is in the \:$x-z$\: plane, the dust vorticity 
$\boldsymbol{\omega} = \nabla\times\textbf{U}$ is directed purely along 
\:$\uvec{y}$\:. In  small Reynolds number 
\:$R_e = \frac{L U}{\mu}$\: limit for the dust fluid, as described in Ref.~{\color{blue}{\cite{M.Laishram}}},  
 nonlinear convective terms are negligible as compared to diffusive terms and 
under this condition Eqs.~{\color{blue}\eqref{1}} and {\color{blue}\eqref{2}} 
combine to produce the equilibrium equation for $\omega$,
%
\begin{equation} \label{9}
	\mu  \nabla^2\omega - \nu\omega + \frac{q_d}{m_d}(\nabla\times\textbf{E}_\textbf{a})_{y} =0.
 \end{equation}
From the Eq.~{\color{blue}\eqref{ambiploar}}, we have;
\begin{equation}\label{10}
\nabla\times\textbf{E}_\textbf{a}=\eta n_e q_e\left(\nabla\times\frac{\textbf{E}\times\textbf{B}}{|\textbf{B}|^2}\right).
\end{equation}
 Using the standard vector identity for the curl of a vector cross product, we write,
\begin{equation}\label{curlExB}
\begin{split}
	\nabla\times\frac{\textbf{E}\times\textbf{B}}{|\textbf{B}|^2} = \textbf E\left(\nabla \cdot\frac{\textbf{B}}{|\textbf{B}|^2}\right)-\frac{\textbf{B}}{|\textbf {B}|^2}\nabla\cdot\textbf{E} + \left(\frac{\textbf{B}}{ |\textbf{B}|^2}\cdot\nabla\right)\textbf{E}\\ -(\textbf{E}\cdot\nabla)\frac{\textbf{B}}{|\textbf{B}|^2}. 
\end{split}
\end{equation} 
We note that the first, second and third term of right hand side of 
Eq.~{\color{blue}\eqref{curlExB}} either vanish or negligible for our setup.  
Specifically, the first term vanishes because magnetic field is 
divergence-free, $\nabla\cdot\textbf{B} = 0$, and varies only along $z$, 
while the second term approaches zero because sheath electric field $E_s$ 
is assumed slowly varying in comparison to variation in $B$ in the dust domain. 
The third term vanishes because there is no gradient in $\textbf{E}$ along 
$\textbf{B}$ either.  
Under these conditions, Eq.~{\color{blue}\eqref{curlExB}} reduces to,
\begin{equation}\label{12}
\nabla\times\frac{\textbf{E}\times\textbf{B}}{|\textbf{B}|^2} = -(\textbf{E}\cdot\nabla)\frac{\textbf{B}}{|\textbf{B}|^2}.
\end{equation}
Using the linearized variation of the magnetic field 
$\textbf{B}(z) = B_0(1+\alpha z)$, and the sheath electric field as
$\textbf{{E}}$ = $E_s\uvec{z}$, the right hand side of the 
Eq.~{\color{blue}\eqref{12}} becomes,
\begin{equation}\label{13}
	\nabla\times\frac{\textbf{E}\times\textbf{B}}{|\textbf{B}|^2} = -E_s\frac{\partial}{\partial z}\frac{1}{B_0(1+\alpha z)}\uvec{y}.
\end{equation}
 By using Eq.~{\color{blue}\eqref{13}} in 
Eq.~{\color{blue}\eqref{10}} we 
finally obtain $\nabla\times\textbf{E}_\textbf{a}$ as,
\begin{equation} \label{14}
	\nabla\times\textbf{E}_\textbf{a}=-\eta n_{e} q_{e}E_s\frac{\partial}{\partial z}\frac{1}{B_0(1+\alpha z)}\uvec{y},
 \end{equation}
 so that the Eq.~{\color{blue}\eqref{9}} can be written as,
%
%
\begin{equation} \label{17}
\mu  \nabla^2\omega - \nu\omega + \kappa\omega_a =0,
 \end{equation}
 where the coefficient kappa $\kappa$ and qunatity $\omega_a$ are respectively 
given as,
\begin{equation} \label{def-kappa}
\kappa = \frac{\eta n_e q_d q_e}{m_d},
 \end{equation}
and, 
%
\begin{equation} \label{driver}
	\omega_a = \omega_{a0}\frac{1}{\left(1+\alpha z\right)^2}
	= \frac{E_{s}}{B_{0}}\frac{\alpha}{\left(1+\alpha z\right)^2}.
 \end{equation}
 Here $\omega_a$ is the strength of the vorticity source provided by $\textbf{E}$ and the magnetic field varying along $z$,

The continuity equation for the incompressible dust fluid {\color{blue}\eqref{continuity}} 
allows one to define the streamfunction $\Psi$ such that 
$\textbf{U}=\nabla\times (\Psi \hat{y})$ prescribing its the relationship with $\omega$ as,
\begin{equation} \label{omega}
  \omega = -\left(\frac{\partial^2\Psi }{\partial z^2} +  \frac{\partial^2\Psi }{\partial x^2}\right)
 \end{equation}
The quantity $\omega_{a}$ replaces the vorticity produced by a 
non-conservative drive, for example, by the ion drag force in 
Ref.~{\color{blue}{\cite{M.Laishram}}}.
Note that a pure electrostatic field produced by sheath structure would still
have a zero vorticity and therefore can not act as a source for the dust 
vortex flow. Remarkably, for a finite number of terms to survive in 
Eq.~({\color{blue}\ref{curlExB}}) it is required that a magnetic field be 
present essentially. In case of 
a nonuniform magnetic field and uniform electric field, the first and 
fourth term can survive. For a magnetic field varying only along $z$ however 
(as in the present case) the first term vanishes but the fourth term still 
provides finite contribution.  A few more interesting cases 
would be as follows.
In the case of uniform magnetic field, on the other hand, 
a finite contribution is still possible from second and third terms if a 
gradient in the electric field is present. If the gradient in the electric 
field is orthogonal to the direction of $B$ the third term vanishes but the 
second term can still be finite. In the present setup, as described above, 
the magnetic field gradient is assumed to be present along a nearly uniform 
electric field and therefore only the fourth term provides a finite 
contribution, for the simplicity of the analysis.

Eq.~{\color{blue}\eqref{17}} is a fourth order partial differential 
equation in stream-function and can be solved under the 
assumptions made in Ref.~{\color{blue}{\cite{M.Laishram}}}, namely, that 
the variation of $\Psi$ is determined by the independent choice of driver 
scale variation along the two orthogonal directions. The confinement 
domain can therefore be elongated such that the shear effect are stronger 
only along one of its dimensions. 
Considering the dependence along $x$ of $\Psi$, in comparison to
that along $z$, to be produced by the variation the $E_{a}$, and that 
along $z$ to be independently prescribed by variation of $B$, we 
choose $L_{x} \gg L_{z}$ and the dependence on $z$ 
and $x$ can be treated via a separable function for $\Psi$. This allows 
$\Psi$ to be expressed in the form of the product 
$\Psi$ = $\Psi_x(x)$ $\Psi_z(z)$ and the equation becomes
\begin{equation} \label{vorticity-equation}
\begin{split}
	\left(\frac{\partial^4\Psi_z }{\partial z^4}+ 2k_x^2\frac{\partial^2\Psi_z }{\partial z^2}- K_1\frac{\partial^2\Psi_z }{\partial z^2}+ \Psi_z\frac{\partial^4 }{\partial x^4}-K_1\Psi_z\frac{\partial^2 }{\partial x^2} \right)\Psi_{x}\\ - K_2 \omega_a = 0
\end{split}
 \end{equation}
having two parameters, $K_1$ = $\frac{\nu}{\mu}$ and 
$K_2$ = $\frac{\kappa}{\mu}$.
The procedure for the solution of Eq.~\ref{vorticity-equation} is described in
the Sec.~\ref{NM1}.
\section{boundary value problem in Cartesian setup}\label{NM1}
Eq.~{\color{blue}\eqref{vorticity-equation}} is treated as an eigenvalue 
equation for $\Psi_{z}$ which is nonzero (bounded) in the region 
$0 < z/L_z <1$. However, since the in-homogeneity is introduced by the driver
term $K_{2}\omega_{a}$ which remains independent of the boundaries imposed on 
the dust fluid, a numerical solution with sufficient number of eigenmodes 
is considered as treated below. 

We represent both, the driven dust stream-function and the driver field 
in term of linear combinations of common eigen functions satisfying the 
boundary condition imposed on $\Psi_{z}$ and write,
$\Psi_z = \sum^{\infty}_{m=1}c_m\phi_m$ and 
$\omega_a = \omega_{xa}\sum^{\infty}_{m=1}b_m\phi_m$ {\color{blue}{\cite{M.Laishram}}} 
with,
$\phi_m$ = $\sin(k_m z)$, where $k_m$ = $\frac{\pi m}{L_z}$.
The Eq.~(\ref{vorticity-equation}) thus takes the form,
\begin{equation}\label{ve}
	\left(F~\sum^{\infty}_{m=1}c_m\phi_m\right)\Psi_{x} = \left(K_2\sum^{\infty}_{m=1}b_m\phi_m\right)\omega_{xa}
\end{equation}
Where $F$ represents the operator,
\begin{equation}\label{operator}
  F = {\ \frac{\partial^4 }{\partial z^4}+ 2k_x^2\frac{\partial^2 }{\partial z^2}- K_1\frac{\partial^2 }{\partial z^2}+ \frac{\partial^4 }{\partial x^4}-K_1\frac{\partial^2 }{\partial x^2}}
\end{equation}
In order to reduce Eq.~(\ref{ve}) to a solvable eigenvalue problem under certain 
special conditions, we now consider the case where along the direction of weak
variation, $x$, the dust is driven as a single eigenmode 
$\Psi_{x}=c_{x}\phi_{x}$, by a corresponding single eigenmode of the driver
$\omega_{xa}=b_{x}\phi_{x}$. 
This readily allows us to redefine the known and unknown coefficients $b_m$ and 
$c_m$, respectively, as 
$b_m\rightarrow b_{x}b_{m}$ and $c_m\rightarrow c_{x}c_{m}$,
and the Eq.~(\ref{ve}) reduces into a solvable form,
\begin{equation}\label{eve}
	F~\sum^{\infty}_{m=1}c_m\phi_m = K_2\sum^{\infty}_{m=1}b_m\phi_m.
\end{equation}
The eigenvalue equation for the operator $F$ can be written as,
\begin{equation}\label{23}
(F-\lambda_m)\sin(k_m z)=0,
\end{equation}
where the eigenvalues $\lambda_m$ are, 
\begin{equation}\label{24}
	\lambda_m = k_m^4 + K_1 k_m^2-(2k_m^2+K_1)k_x^2+k_x^4.
\end{equation}
The unknown coefficients $c_{m}$ need to be obtained from the solution of 
a set of $M$ simultaneous equations to be given from the relation,
%
\begin{equation}\label{set}
\displaystyle\sum_{m=1}^{M}(\lambda_m c_m - K_2 b_m)\sin(k_m z)=0 ,
\end{equation}
where $M$ represents the limiting value of the number of modes sufficient 
to reproduce the smallest length scale in the problem. When rewritten in terms 
of unknown coefficients $c_{m}$ of the problem which need to be determined,
(\ref{set}) takes a more familiar form,
\begin{equation}\label{set1}
	A_{im} c_{m}  = B_{i},
\end{equation}
 where $A_{ij}$ is the matrix of the coefficients, with $i$ being the index 
for the 
spatial locations $z_{i}$ where solutions values ($\Psi_{z}(z_{i})$ values) 
are desired. These coefficients and the vector $B_{i}$ in this new form 
become,
\begin{eqnarray}\label{def-A}
	A_{im} &=& \lambda_m\sin(k_m z_i) 
\\\label{def-B}
{\rm and}~~~~~~B_i &=& K_2 \displaystyle\sum_{j=1}^{M} b_j\sin(k_j z_i)
\end{eqnarray}
Since the elements of $\textbf{A}$ and $\textbf{B}$ are known,  
the solution for the streamfunction involves determining the values of 
coefficients $c_{m}$ using the inversion,
\begin{equation}\label{inversion}
c_m = A^{-1} B
\end{equation}
The boundary conditions on $U_x= -\frac{\partial \Psi_z}{\partial z}$
can further be imposed requiring the $z$ profile of streamfunction to
have a desired derivative. This procedure is adopted for the solutions 
presented in the Sec.~\ref{NM2} where effect of magnetization of electrons 
are investigated on the dust vortex flow dynamics.
\section{DUST VORTEX FLOW SOLUTIONS IN A NON-UNIFORM MAGNETIC FIELD}\label{NM2}
In the following analysis, the values of quantities (mass, length, velocity 
etc.) associated with a typical dusty plasma set up are used to scale the 
parameters and variables involved in the above analytic formulation.
We accordingly use dust mass $m_d$,  
ion acoustic velocity $U_{A}$, 
dust charge $q_d$ and 
length of the simulation box $L_z$ values in a typical dusty plasma to 
normalize our variables such that, 
the variables
 $\omega$, 
 $\kappa$ and
 $\nu$
 have the unit
 ${\frac{U_{A}} {L_z}}$, 
 the variables
 $\Psi$ and
 $\mu$
 have the unit
 ${U_{A} L_z}$
 and the variables
 $E_s$, 
 $\eta$ 
 and $B_0$ 
 have the units 
 ${\frac{m_d U_{A}^2}{q_d L_z}}$, 
 ${\frac{m_d U_{A} L^2_z}{q^2_d }}$, 
 and  ${\frac{m_d U_{A}}{q_d L_z}}$, 
 respectively.

For parameters corresponding to a typical weakly magnetized dusty plasma {\color{blue}{\cite{Melzer}}} 
where $m_d$ = $1\times {10^{-14}}$ Kg,  $L_z$ =$0.1$ m, 
$U_{A}$ = $2.5 \times {10^{3}}$ m/sec, charge on the dust $q_d$ = 
$1.6\times{10^{-16}C}${\color{blue}{\cite{Khrapak2}}}, electric field $10^{3}$ Vm$^{-1}$ and magnetic 
field 10-100 G, we estimate the
typical values of our input parameters as 
the electric field $E_s$ = 2.5$\times{10^{-7}}$ ${\frac{m_d U_{A}^2}{q_d L_z}}$, 
the magnetic field $B_0$ = 6.4$\times{10^{-10}}$ 
${\frac{m_d U_{A}}{q_d L_z}}$, and $\alpha =1 $ $L_{z}^{-1}$
which provide,  
$\omega_{a0}$ = 4$\times{10^{2}}$ $U_{A}$/$L_z$.
$\eta$ = $  {10^{-20}}$ ${\frac{m_d U_{A} L^2_z}{q^2_d }}$ {\color{blue}{\cite{FF}}}, 
$n_e$ =  $ {10^{13}}$ $L_{z}^{-3}$, 
and $\kappa$ = ${10^{-10}}$ ${\frac{U_{A}} {L_z}}$.

Note that a rather stronger resistivity because of the electron-neutral 
(e-n) collisions 
would be more appropriate for high pressure cases experimentally analyzed 
by Puttscher and Melzer {\color{blue}{\cite{Melzer}}} where the $\eta\equiv \eta_{en}$ value should 
be a few order higher than 
the standard Spitzer resistivity mentioned above. Considering that the 
$E_{a}$ value produced rather by electron-neutral resistivity 
($\sim \eta_{en}n_e \frac{E_s}{B_{0}}$) still remains sufficiently lower 
than the lowest order electric field (or above $E_s$ value), as required 
for the analysis to hold good, larger values of $\eta$ remain equally 
admissible.
We however use the above representative Spitzer resistivity value for all 
our computations presented here.
For the present numerical solutions we have used large number of eigenmodes 
(M=200) to express the resulting dust stream-function $\Psi_z$ at equal 
number of locations $z_{i}$.

The solutions in the present treatment are obtained in the rectangular 
domain measuring $L_{z}$ along $z$ and $20 L_{z}$ along $x$ directions
as appropriate for the limit $k_{x}\ll k_{z}$ of the analysis.
The profile for the source field $\omega_{a}$ given by Eq.~(\ref{driver}) 
considered for this analysis, using $B_{0}= 6.4\times{10^{-10}}$ ${\frac{m_d U_{A}}{q_d L_z}}$ and $\alpha=1$ $L_{z}^{-1}$, is presented 
in Fig.{\color{blue}\ref{2dprofiles}(a)} as 
generated by a linear variation in the applied magnetic field that remains 
independent of the dust boundaries. The choice of applying boundary condition 
for the dust flow is available only at $z=0$ and $z=L_{z}$ which correspond 
to two adjacent sides of the rectangular dust confinement domain in the 
$x$-$z$ plane. We have applied dissimilar boundary conditions for the 
dust flow velocity along these two boundaries. For example, a no-slip 
boundary condition is applied at the lower boundary, $z=L_{z}$ while no
control on the velocity values is done on the boundary at $z=0$ where
the dust velocity is freely determined by the driver strength.

\begin{figure}[!h]
 \centering
 \includegraphics[width=95mm]{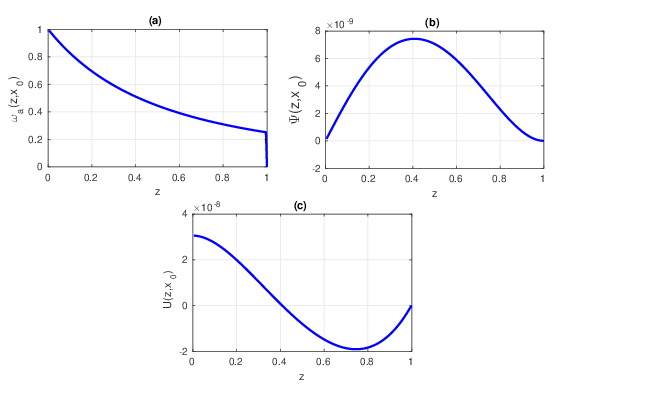}
 \caption{Functions for dust flow profile plotted at $x$ = $x_0$ = $L_x$/2. (a) 
driver profile, $\omega_a$ = 4$\times{10^{2}}$, (b) dust stream-function, 
$\Psi(z,x_0)$, and (c) x-component of dust velocity profile for $\mu$ = 0.01 ${U_{A} L_z}$, 
$\nu$ = 0.1 ${\frac{U_{A}} {L_z}}$, $B_{0}= 6.4\times{10^{-10}}$ ${\frac{m_d U_{A}}{q_d L_z}}$, $\alpha=1$ $L_{z}^{-1}$ and $\kappa$ = ${10^{-10}}$ ${\frac{U_{A}} {L_z}}$.} 
 \label{2dprofiles}
\end{figure}

A range for value of dust viscosity $\mu$ = 0.01 to 0.001 ${U_{A} L_z}$, 
is chosen 
considering the dust fluid flow to be in small Reynolds number limit 
($\le$ 1) as given in Sec.~\ref{NM} which is consistent with the present 
linear limit considered of the model. 
The collision frequency $\nu$ = 0.1 $U_{A}$/$L_z$ is considered here for 
sufficient high pressure regime, for example that described by Puttscher and Melzer {\color{blue}{\cite{Melzer}}} where ambipolar field effects dominates
the ion drag force.

The profile for dust streamfunction $\Psi$ at $x = L_{x}/2$ is 
presented in Fig.~{\color{blue}\ref{2dprofiles}(b)} driven by source field 
$\omega_a$ 
which is provided by the combination of sheath electric field and non-uniform 
applied magnetic field. The boundary conditions on dust streamfunction 
discussed above ensures zero dust velocity at $z = L_{z}$, as plotted in 
Fig.{\color{blue}\ref{2dprofiles}(c)}, and is independent of the driver 
strength at this 
boundary.  
A zero net flux of dust particles crosses the $x = L_{x}/2$ line
as indicated by the velocity profile presented in 
Fig.{\color{blue}\ref{2dprofiles}(c)} which has equal area under the 
positive and negative region of the curve considering that the 
dust fluid is incompressible. 

The ambient electric field along $z$ (0,0,\textbf{E}) and the applied weak 
magnetic field along $y$ (0,\textbf{B},0) cause only the electrons to drift 
in  negative x-direction as the ions are unmagnetized. The displacement of 
electrons generates a space-charge field which is directed along negative $x$ 
direction. The negatively charged dust must flow along positive $x$ because 
of this electric field and it 
experiences a force ($\textbf{F}_{E}$=-$q_{d}\textbf{E}_\textbf{a}$). 
However since the 
magnetic field has a gradient along $z$, the electron drift is nonuniform 
in $z$ and the space charge field $\textbf{E}_\textbf{a}$ generated 
by the electrons displacement is nonuniform along $z$. 
As a result, the dust velocity profile has a change of sign in the region as 
the dust experiences a larger force in positive $x$ direction at small 
$z$ values and must flow in along this force. Due to its incompressible 
character, however, a return flow is set up through the region of large 
$z$ where the ambipolar field $\textbf{E}_\textbf{a}$ is weaker and therefore 
the dust velocity sign is opposite, setting up vortex flow.

 \begin{figure}[!h]
   \centering
 \includegraphics[width=100mm]{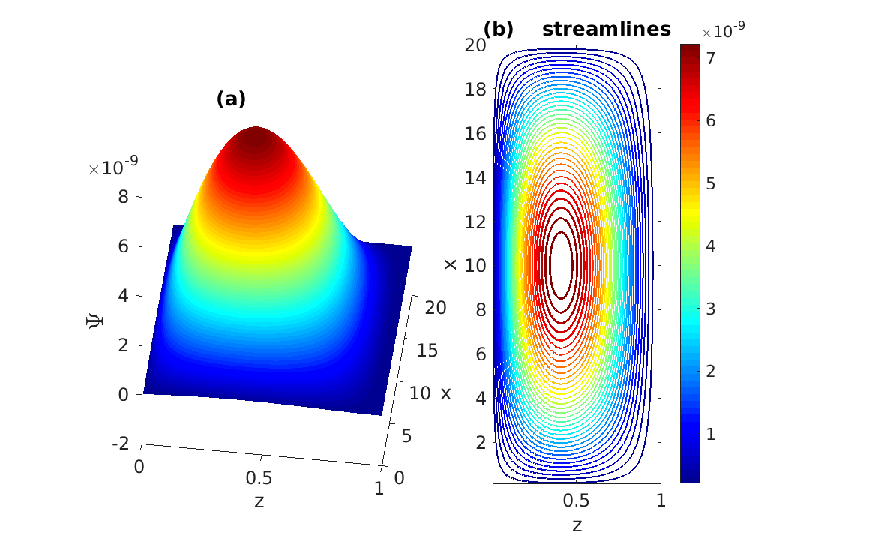}
  \caption{2D-functions for  dust flow profiles, (a) dust stream function $\Psi(z,x)$, and (b) dust streamlines.}  
   \label{streamlines}
\end{figure}

The sign of the dust flow in the region of strong ambipolar field in our case 
is consistent to the Melzer and Puttscher {\color{blue}{\cite{Melzer3}}} where they 
have observed the displacement of the dust particles in negative 
$\textbf{E}\times\textbf{B}$ direction at sufficiently high gas pressure.

\begin{figure}[!h]
    \centering
    \includegraphics[width=95mm]{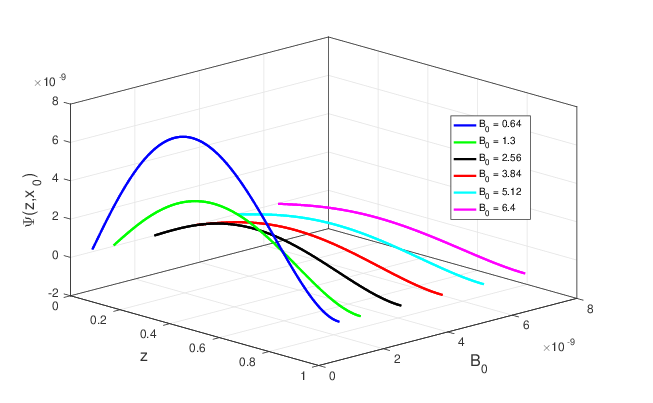}
  \caption{ Dust stream-function profiles with different values of magnetic field $B_{0}$ and {\color{red}{$\alpha$} = 1}.}
    \label{streamfunction-B}
\end{figure}

The 2D surface plot of the streamfunction solution is presented in 
Fig.{\color{blue}\ref{streamlines}(a)} for the case using 
$\mu= 0.01$ ${U_{A} L_z}$, $\nu=0.1$ ${\frac{U_{A}} {L_z}}$, $\alpha=1$ $L_{z}^{-1}$, $\kappa= {10^{-10}}$ ${\frac{U_{A}} {L_z}}$, which obeys the 
applied boundary conditions at the boundaries at $z=0$ and $z=L_{z}$, 
respectively. 
Similarly, the confinement of dust fluid in x-direction is ensured by 
uniformity in the value of $\Psi$ along z at boundaries x = 0, $L_{x}$, 
such that $\partial \Psi/\partial z=0$ at these boundaries and the dust 
fluid remains confined in region $0 < x < L_{x}$. 
The topology of the surface plot corresponds to a vortex 
structure in dust velocity field and the corresponding streamlines of dust 
flow is presented in Fig.{\color{blue}\ref{streamlines}(b)}.
This shows that the ambipolar electric field can act as a source of finite 
vorticity in the dust flow dynamics. Although the ambient time independent 
electric and magnetic fields included here provide only conservative forcing 
mechanisms, when a drift based mechanism is considered the dust vortex flow 
solutions are recovered. The dust streamlines in 
Fig.~{\color{blue}\ref{streamlines}(b)} is clear evidence of dust vortex formation. 
The emergence of macroscopic dust vortex flow becomes possible by 
non-zero value of parameter $\alpha$ responsible for the simplest 
non-uniformity introduced by $\alpha$.

\begin{figure}[!h]
    \centering
   \includegraphics[width=95mm]{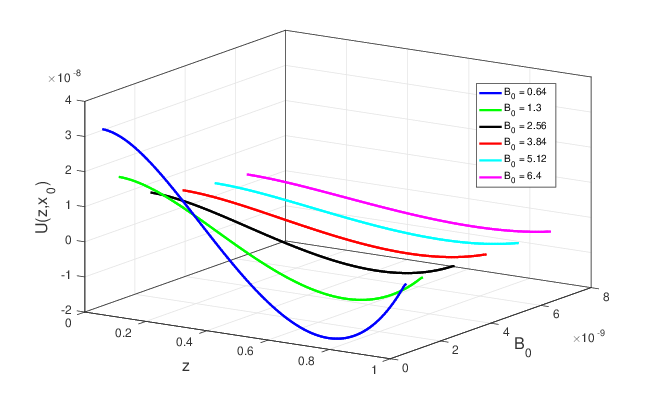}
\caption{x-component of dust velocity profiles with different values of magnetic field} $B_{0}$ and {\color{red}{$\alpha$} = 1}.
    \label{velocity}
\end{figure}

The dust streamfunction profiles plotted at $x=L_{x}/2$ with the variation of 
magnetic field strength $B_0$ is presented in 
Fig.~{\color{blue}\ref{streamfunction-B}}, for the parameters 
$\mu= 0.01$ ${U_{A} L_z}$, $\nu=0.1$ ${\frac{U_{A}} {L_z}}$, $\alpha=1$ $L_{z}^{-1}$, $\kappa= {10^{-10}}$ ${\frac{U_{A}} {L_z}}$. 
Dust gradient of streamfunction gradually reduces to a minimum value with 
increase in applied magnetic field strength $B_0$. Therefore, as from the 
Eq.~(\ref{omega}), vorticity associated with circulation motion of dust flow 
field also tends to reduced with strength of applied magnetic field $B_0$. 
The corresponding dust velocity flow field profiles are presented 
in Fig.{\color{blue}\ref{velocity}} showing that the magnitude of the 
maximum dust velocity achieved at $z=0$ decreases with increase in 
applied magnetic field strength $B_0$. 
In present analysis the dust velocity is determined by the combination of 
ambipolar electric field, $\textbf{E}_\textbf{a}$, neutral drag, 
$\nu$ and kinematic viscosity, $\mu$, of dust fluid. For the fixed values 
of $\nu$ and $\mu$, however, the dust velocity is determined only by ambipolar 
force as presented in Fig.{\color{blue}\ref{velocity}}. 
Since the ambipolar electric field 
$\textbf{E}_\textbf{a}$ has a inverse relation with magnetic field as given by 
Eq.({\color{blue}\ref{ambiploar}}), the electrostatic force 
($\textbf{F}_{E}$=-$q_{d} \textbf{E}_\textbf{a}$) arising from the ambipolar 
field reduces with the magnetic field. As a result, the dust vortex flow 
weakens at higher magnetic field as shown in Fig.{\color{blue}\ref{velocity}}. 
The effect of $\textbf{B}$ field strength analyzed on the ambipolar field 
here is therefore in contrast to the effect of $\textbf{E}$ field strength 
that may be present in the plasma sheath and would instead strengthen the 
vortex flow.
 
\begin{figure}[!h]
 \centering
  \includegraphics[width=95mm]{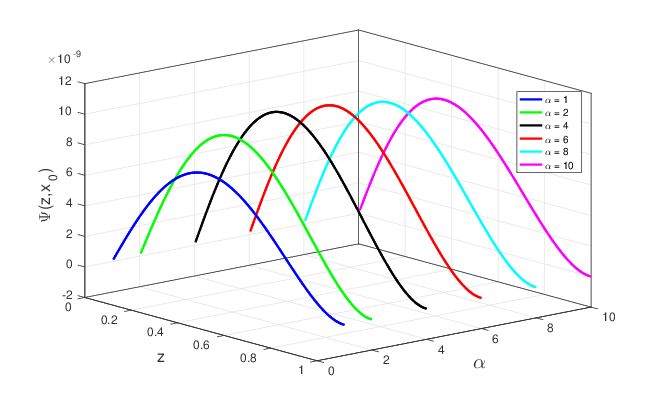}
 \caption{ Dust stream-function profiles with different values of $\alpha$ and constant magnetic field  {\color{red}{$B_0$} = 6.4$\times{10^{-10}}$}.}
  \label{streamfunction-alpha}
\end{figure}

\begin{figure}[!h]
    \centering
    \includegraphics[width=95mm]{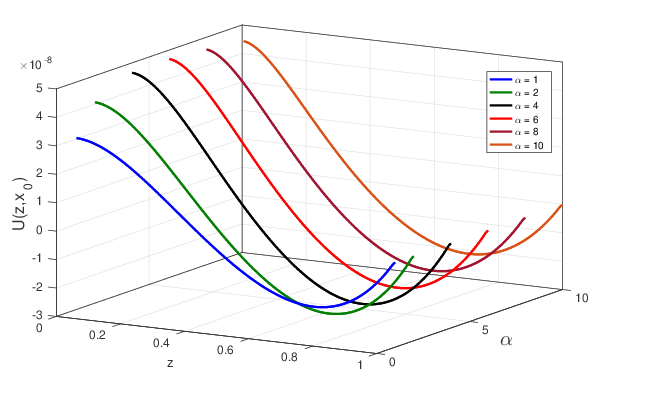}
\caption{x-component of dust velocity profiles with different values of $\alpha$ and constant  magnetic field}  {\color{red}{$B_0$} = 6.4$\times{10^{-10}}$} .
  \label{velocity-alpha}
\end{figure}

Analysis of dust flow field with varying applied magnetic field gradient 
$\alpha$ is presented in Figs.{\color{blue}\ref{streamfunction-alpha}} and 
{\color{blue}\ref{velocity-alpha}}. 
Dust streamfunction peak value first increases and then slightly decreases with 
increasing $\alpha$ as shown in Fig.{\color{blue}\ref{streamfunction-alpha}} 
at constant magnetic field strength $B_0(z=0)$ = 6.4$\times{10^{-10}}$ ${\frac{m_d U_{A}}{q_d L_z}}$. 
The corresponding dust velocity field profiles are presented in 
Fig.{\color{blue}\ref{velocity-alpha}}. Peak dust velocity (at $z=0$) 
similarly shows a maximum with respect to $\alpha$ however its variation 
remains comparatively weaker at larger values of $\alpha$. 

 \begin{figure}[!h]
  \centering
  \includegraphics[width=95mm]{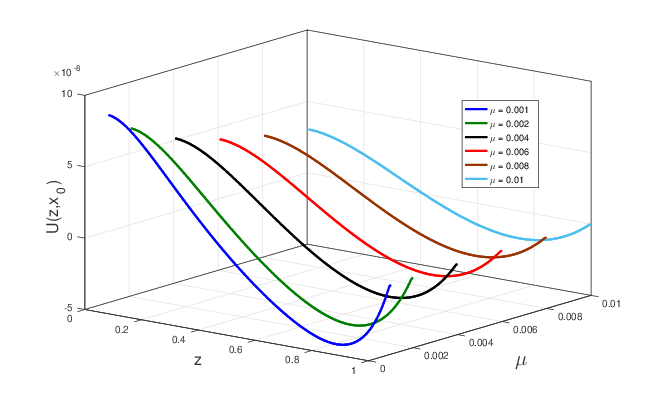}
 \caption{x-component of dust velocity profiles with different values of viscosity of dust fluid} $\mu$,  for {\color{red}{$\alpha$} = 1} and {\color{red}{$B_0$} = 6.4$\times{10^{-10}}$} .
  \label{velocity-mu}
\end{figure}

Effect of kinematic viscosity $\mu$ on dust velocity in non-uniform 
magnetic field has been analyzed in Fig.{\color{blue}\ref{velocity-mu}} 
with magnetic field strength $B_0(z=0)$ = 6.4$\times{10^{-10}}$ ${\frac{m_d U_{A}}{q_d L_z}}$
and $\alpha$ = 1$L_{z}^{-1}$. With the application of the no-slip boundary condition 
at the boundary $z=L_{z}$ we note in Figs.~\ref{velocity} and 
\ref{velocity-alpha} a negligible effect of magnetic filed 
strength and its gradient, respectively, on the width of the boundary layer
that forms and must shrink with reducing $\mu$ {\color{blue}{\cite{M.Laishram}}} as presented in 
Fig.~\ref{velocity-mu} for the present case. The vortex flow is also seen to 
weaken with increasing $\mu$.
\section{Dust vortex solutions for multipolar form of the ambipolar field}\label{MSD}
We finally explore the cases where the modulation in magnetic field strength 
can result in multipolar structure of the ambipolar electric field 
${\bf E}_{a}$. This effect is achieved by examining cases with individual 
modes in the magnetic field gradient driving the vortex flow and using 
increasing values $m=1,2$ and 3 of the mode number $m$ while using the 
strength of this effect as determined by the factor $\kappa$. 
 \begin{figure}[!h]
    \centering
    \includegraphics[width=90mm]{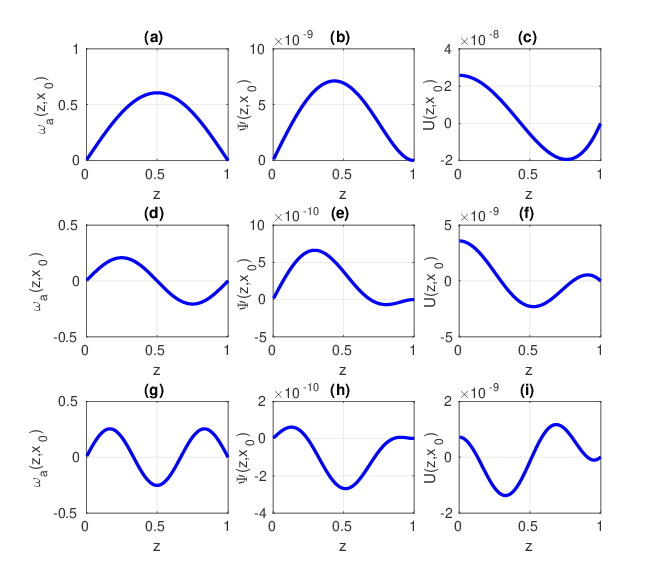}
   \caption{ Functions for the dust flow profiles with different driver mode number (a) driver profile $\omega_a$ (b) dust stream function $\Psi(z,x_0)$ and (c) x-component of dust velocity profile for mode (m=1) and similarly fig. (d),(e),(f) and (g),(h),(f) for m = 2 and m =3 , respectively, for  {\color{red}{$\alpha$} = 1} and {\color{red}{$B_0$} = 6.4$\times{10^{-10}}$.}}
 \label{multiple driver}
 \end{figure}
Note that the quantity $\omega_{a}$ is an effective source of vorticity produced by 
the shear in the ambipolar electric field present in 
Eq.~(\ref{vorticity-equation}) and arising, in this case, from a rather 
{\em wave-like} spatial variation of the ambient magnetic field.

\begin{figure}[!h]
    \centering
    \includegraphics[width=95mm]{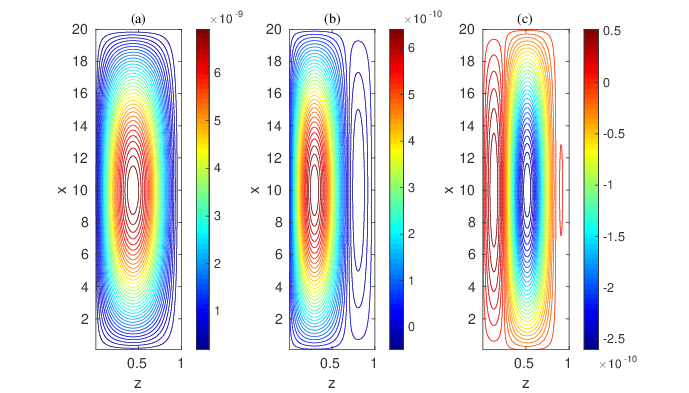}
    \caption{ streamlines for the dust fluid flow, with different driver mode number (a) m=1 (b) m=2 (c) m=3 using parameters  {\color{red}{$\alpha$} = 1}  and {\color{red}{$B_0$} = 6.4$\times{10^{-10}}$.}}
 \label{multiple streamline}
 \end{figure}

As presented in Fig.{\color{blue}\ref{multiple driver}},
the source field having an individual mode number $m = 1$ in
Fig.{\color{blue}\ref{multiple driver}}(a) produces a single peak profile of 
the streamfunction plotted in Fig.{\color{blue}\ref{multiple driver}}(b) 
and the corresponding dust velocity plotted in 
Fig.{\color{blue}\ref{multiple driver}}(c) has a single dust vortex.
The streamlines corresponding to this case are plotted in 
Fig.{\color{blue}\ref{multiple streamline}}(a) showing a single 2D vortex 
flow structure.
For the case $m=2$ as presented in 
Fig.{\color{blue}\ref{multiple driver}}(d)-(f), however, a velocity profile 
consistent with a set of two counter-rotating vortices emerges. The 
corresponding dust streamlines in 
Fig.{\color{blue}\ref{multiple streamline}}(b) clearly show this set of 
counter-rotating vortex flow structures. A further increase in the $m$ 
value, $m=3$, similarly produces a sequence of three counter-rotating dust 
vortex flow structures as presented in 
Fig.{\color{blue}\ref{multiple driver}}(g)-(i) and 
Fig.{\color{blue}\ref{multiple streamline}}(a). The third vortex close 
to the boundary $z=L_{z}$ in this case however has a very low strength 
because of the flow satisfying a no-slip boundary condition at this 
boundary.
 \section{SUMMARY AND CONCLUSIONS}\label{SC}
  To summarize, we have presented an analysis of a dust vortex flow in 
the electrically charged dust medium suspended in a weakly magnetized plasma. 
We have examined the cases where a sheared $\textbf{E}\times\textbf{B}$ drift, 
arising from a spatially non-uniform magnetic field, is able to drive a vortex 
motion of the suspended dust fluid. By employing the 
$\textbf{E}\times\textbf{B}$ effect on dust particles as recovered and 
described by Melzer and Puttscher {\color{blue}{\cite{Melzer3}}}, we have shown that
the ambipolar electric field
can act as a source of finite vorticity in the dust flow dynamics. The 
expressions derived by us use the circulation of the ambipolar field 
generated by 
$\textbf{E}\times\textbf{B}$ drift of the electrons as a driver for
vortex flow of the dust motion allowing study of its behavior in presence of 
nonuniform magnetic field. The results characterize nature of dependence of 
the dust vortex motion on the strength of magnetic field and its gradient.

The dust streamfunction 
solutions in a Cartesian setup obtained under applied non-uniform magnetic 
field $B(z)$ and its linear gradient $\alpha$, over the dust confinement domain,
show that a combination of conservative fields (magnetic and electric field) 
can generate a finite circulation in dust flow field. 
The resulting dust vortex flow driven is therefore driven in the absence of 
any non-conservative fields, e.g., friction, ion drag and the dust 
charge variation. A multipolar nature of the ambipolar electric field
is additionally recovered for wave-like nature of the spatial gradients 
and is examined in terms of a sequence of 
counter-circulating dust vortex flow produced by it for larger mode number 
of the magnetic field variation. 
The vortex flow motion of the highly charged dust medium in a magnetized 
plasma environment, arising purely from the field non-uniformity can be 
an interesting effect for magnetized dusty plasma, both laboratory experiments  
and in natural conditions, such as in astrophysical circumstances.
The present first study of this process can thus provide quantitative inputs 
for conducting the related laboratory experiments for exploring the deeper 
correlation between the two. 


 \section{AIP PUBLISHING DATA SHARING POLICY}\label{SC}
 The data that support the findings of this study are available from the corresponding author upon reasonable request.
 

 \bibliographystyle{apsrev4-1}
 \bibliography{paper}


\end{document}